\documentstyle[12pt]{article}
\newcommand{\be}{\begin{equation}}
\newcommand{\ee}{\end{equation}}
\newcommand{\bear}{\begin{eqnarray}}
\newcommand{\ear}{\end{eqnarray}}
\newcommand{\yc}{y_{\rm cut}}
\newcommand{\Nc}{N_{\rm cut}}
\newcommand{\Nt}{N_{\rm tot}}
\newcommand{\bm}[1]{\mbox{\bf #1}}
\renewcommand{\vec}{\bm}

\begin{document}
\begin{titlepage}
\begin{flushright}
HD--THEP--94--43
\end{flushright}\quad\\
\vspace{-17mm}
\begin{flushright}
PITHA--94--58\hspace*{6.8mm}
\end{flushright}
\vspace{1.8cm}
\begin{center}
{\bf\LARGE CP VIOLATION AND THE WIDTH $Z\rightarrow b\bar{b}$}\\
\vspace{1cm}
W.\ Bernreuther\\
\bigskip
Institut  f\"ur Theoretische Physik\\
Physikzentrum RWTH\\
Sommerfeldstr.~26--28, D-52056 Aachen, FRG\\
\vspace{1cm}
and\\
\vspace{1cm}
G.\ W.\ Botz, D.\ Bru\ss , P.\ Haberl, and O.\ Nachtmann\\
\bigskip
Institut  f\"ur Theoretische Physik\\
Universit\"at Heidelberg\\
Philosophenweg 16, D-69120 Heidelberg, FRG\footnote[0]{
Submitted for publication to Zeitschrift f\"ur Physik C}\\
\vspace{2cm}
{\bf Abstract:}\\[3mm]
\parbox[t]{\textwidth}{
We discuss the effect of CP--violating $Zb\bar b$, $Zb\bar bG$
and $Zb\bar b\gamma$ couplings on the width
$\Gamma(Z\rightarrow b\bar bX)$. The presence of such couplings
leads in a natural way to an increase of this width relative to
the prediction of the standard model. Various strategies of a
direct search for such CP--violating couplings by using CP--odd
observables are outlined. The number of $Z$ bosons required to
obtain significant information on the couplings in this way is
well within the reach of present LEP experiments.}
\end{center}
\end{titlepage}
\newpage
\setcounter{equation}{0}\noindent
In a series of papers we have investigated possible new
CP--violating couplings which may manifest themselves in $Z$ boson
decays, especially to $\tau$ leptons and to $b$ quarks
[1--6].
There is related work
by other authors \cite{7}. Experimental results on the CP--violating
weak dipole moment of the $\tau$ lepton were published in \cite{8}.
In this note we want to update and extend the analyses presented in
\cite{1,3} for the decays of the $Z$ boson into bottom quarks. This
seems timely because in recent years large samples of $b\bar{b}$
events have been selected by LEP experiments using microvertex
detectors.
\par
Important information on the strength of new couplings involving
$b$ quarks can be derived from the width $Z\rightarrow b\bar{b}X$.
Its recent LEP average is \cite{9}
\be\label{1}
R^{\rm exp}_b\equiv \frac{\Gamma(Z\rightarrow b\bar{b}X)}
{\Gamma(Z\rightarrow {\rm hadrons})}=0.2197\pm 0.0020\;.
\ee
This average includes the latest OPAL result \cite{9a} of
$R_b=0.2171\pm 0.0021\pm 0.0021$ obtained with a double tagging
method. With a top quark mass of $m_t=175\pm 20$ GeV \cite{10} the
standard model (SM) value
of $R_b$ is \cite{10a}
\be\label{2}
R^{\rm SM}_b= 0.216\pm 0.001\;.
\ee
Hence there is a margin for new physics effects in
$Z\rightarrow b\bar{b}X$.
\par
Of course, even if $R^{\rm exp}_b$ agreed perfectly with the SM
prediction new physics effects in $Z\rightarrow b\bar{b}X$ were
not excluded. Such effects may cancel in the total decay rate
$\Gamma(Z\rightarrow b\bar{b}X)$ but could show up in differential
distributions. We will make some further comments on such a scenario
below.
\par
An interesting possibility is that the channels
$Z\rightarrow b\bar{b}X$ are affected by CP--violating couplings
such that the total amplitude $\cal M$ is the sum of the SM and
a CP--violating one:
\be\label{3}
{\cal M}={\cal M}_{\rm SM}+{\cal M}_{\rm CP} \;.
\ee
In this case the absolute squares of the amplitudes add incoherently
in the width:
\be\label{4}
\Gamma(Z\rightarrow b\bar{b}X)\propto
|{\cal M}_{\rm SM}|^2+|{\cal M}_{\rm CP}|^2 \;,
\ee
thus leading naturally to an enhancement of the width. In this paper
we discuss the magnitude of CP--violating effects one can expect in
$\Gamma(Z\rightarrow b\bar{b}X)$ if indeed the small discrepancy
between the experimental and the SM values for the width
in (\ref{1}), (\ref{2}) is due to a CP--violating amplitude.
Using \cite{10b}
\be\label{5}
\Gamma(Z\rightarrow {\rm hadrons})=1742 \;{\rm MeV} \;,
\ee
we obtain at the 1 s.d.\ level for a possible anomalous term
$\Delta\Gamma$ in the width for $Z\rightarrow b\bar{b}X$:
\bear\label{6}
\Delta\Gamma_{\rm exp}(Z\rightarrow b\bar{b}X)&\leq&
\Gamma_{\rm exp}(Z\rightarrow b\bar{b}X)\mid_{\rm mean\;\,value}
-\Gamma_{\rm SM}(Z\rightarrow b\bar{b}X) \nonumber \\
&&+\left[ \delta\Gamma_{\rm exp}^2{\rm (1\;s.d.)}+
 \delta\Gamma_{\rm SM}^2{\rm (1\; s.d.)} \right]^{1/2} \\
&=& 0.0059\cdot \Gamma(Z\rightarrow {\rm hadrons})
=10.3\;{\rm MeV}\;. \nonumber
\ear
\par
A systematic analysis of CP--violating couplings corresponding to
operators with dimension $\leq$ 6 (\underline{after} symmetry
breaking) relevant for $Z$ decays was given in eq.\ (4.1) of
\cite{1}, where we wrote down the effective Lagrangian. For the
decay $Z\rightarrow b\bar{b}X$ we have in this framework 7
CP--violating coupling constants: the electric,
weak and chromoelectric
dipole moments of the $b$ ($d_b$, $\tilde{d}_b$, $d^\prime_b$),
the parameters $f_{Vb}$, $f_{Ab}$ of the $Zbb\gamma$ vertex
and the parameters $h_{Vb}$, $h_{Ab}$ of the $ZbbG$ vertex.
The dipole moments induce chirality--changing couplings and have
dimension (mass$)^{-1}$. The couplings with $f_{Vb}$, $f_{Ab}$,
$h_{Vb}$, $h_{Ab}$ conserve chirality and these parameters have
dimension (mass$)^{-2}$.
\par
If we calculate the contribution to the width
$\Gamma(Z\rightarrow b\bar{b}X)$ from the CP--odd couplings with
these parameters $d_b,\,\ldots,\,h_{Ab}$, we get for $m_b\neq 0$ a
rather lengthy expression containing also various interference terms.
There, the contribution from the electric dipole moment of the $b$
is proportional to $e^2|d_b|^2$ and thus of the same order as the
electromagnetic radiative correction to the contribution from the
weak dipole term $\tilde{d}_b$. Similarly, the contribution from
the chromoelectric dipole moment $d'_b$ is of order $e^2|d'_b|^2$
and we can give an argument that it
should be discussed together with the QCD radiative correction
to the weak dipole contribution. We plan to present all this in
detail in a future publication.
\par
In this paper we will adopt the following simple procedure: We will
set the quark mass $m_b$ to zero and neglect all terms of order $e^2$
times new couplings $d_b$, $\tilde{d}_b$, $d^\prime_b$, $f_{Vb}$,
$f_{Ab}$, $h_{Vb}$, $h_{Ab}$ in the width. In this approximation the
electric
and chromoelectric dipole terms do not contribute. Then we calculate
the contribution $\Delta\Gamma(Z\rightarrow b\bar{b}X)$ to the width
assuming that either $\tilde{d}_b$ or $f_{Vb}$, $f_{Ab}$ or $h_{Vb}$,
$h_{Ab}$ are different from zero. The result is shown in Table 1.
Here and in the following we use the dimensionless parameters
$\hat{f}_{Vb}$, $\hat{f}_{Ab}$, $\hat{h}_{Vb}$, $\hat{h}_{Ab}$
as defined in (5.1) of \cite{1}:
\par
\hfill\parbox{13.4cm}{
\begin{eqnarray*}
f_{Vb/Ab} & = & -\frac{e^2Q_b}{\sin\vartheta_W\cos\vartheta_Wm_Z^2}
\hat{f}_{Vb/Ab}\;, \\ h_{Vb/Ab} & = & \hphantom{-}
\frac{eg_s}{\sin\vartheta_W\cos\vartheta_Wm_Z^2} \hat{h}_{Vb/Ab}\;.
\end{eqnarray*} }
\hfill\parbox{0.8cm}{\bear\label{61}\ear }
\par\noindent
The result for $\Delta\Gamma$ from the weak dipole term $\tilde{d}_b$
was already given in \cite{1,3}. In the numerical calculations we use
for the fine structure constant at the $Z$ mass $\alpha$=1/129,
$m_Z$=91.19 GeV, $\sin^2\vartheta_W$=0.23, $\alpha_s$=0.12, and
\be\label{7}
\Gamma_{\nu_e\bar{\nu}_e}=\frac{\alpha m_Z}
{24\sin^2\vartheta_W\cos^2\vartheta_WO{}=166\;{\rm MeV}\;.}
\ee
{}From Table 1 we deduce limits on our CP--violating couplings using as
input $\Delta\Gamma_{\rm exp}$ from (\ref{6}):
\par
\hfill\parbox{13.4cm}{
\begin{eqnarray*}
|\tilde{d}_b| \quad&\leq& 3.8\cdot 10^{-17} {\rm e\, cm} \;,\\
\left[(\hat{f}_{Vb})^2+(\hat{f}_{Ab})^2\right]^{1/2}&\leq& 27.5 \;,\\
\left[(\hat{h}_{Vb})^2+(\hat{h}_{Ab})^2\right]^{1/2}&\leq& 2.0 \;.
\end{eqnarray*} }
\hfill\parbox{0.8cm}{\bear\label{8}\ear }
\par
We see that the measured width for $Z\rightarrow b\bar{b}X$ allows
in principle quite sizeable CP--violating couplings for the $b$
quarks. In $Z$ production at electron--positron colliders such
couplings can be searched for directly using the methods of
[1--6].
Here we add
a few comments and present some new calculations.
\par
With unpolarized $e^-$ and $e^+$ beams the initial state is
CP--symmetric, which is the case relevant for LEP. At SLC with
polarized electrons and unpolarized positrons the initial state
is \underline{not} CP--symmetric. Nevertheless, in leading order
one can consider the reaction as a two-step process
\be\label{9}
e^+e^-\longrightarrow Z \longrightarrow b\bar{b}X \;.
\ee
The $Z$ in its rest system is a CP eigenstate (cf.\ \cite{1}). To
leading order CP--violating correlations in the final state are then
also an indicator of CP violation in $Z$ decays. Some calculations
for this case were recently presented in \cite{11}. But now radiative
corrections for polarized $e^+e^-$ collisions have to be carefully
examined for their possibility of faking CP violation. In the
following we will, therefore, concentrate on the case of $e^+e^-$
collisions with unpolarized beams.
\par
In the decay of the $Z$ boson into a pair of $b$ quarks fragmenting
subsequently into $B$ hadrons various final states can occur:
two jets, three jets, two jets plus photon etc. We shall now discuss
these cases with respect to CP violation.
\par
The angular distribution of the jets in two jet decays of the $Z$
does not carry any information on CP violation, as shown in
\cite{1}. To do useful CP studies with the decay
$Z\rightarrow b\bar{b}\rightarrow$2 jets
one has to analyse $b$--$\bar b$ spin and/or spin--momentum
correlations. Thus one needs a spin analyzer for $b$ and $\bar b$.
Parity--odd correlations of the fragmentation products of the
$b$, $\bar b$ (the``handedness'' of the $b$ and $\bar b$ jets) may
provide such spin analyzers
[15--18].
But the relevant analyses
have just started to be done \cite{16}. Thus, at the moment we cannot
obtain any information on CP--violating couplings from 2 jet events.
\par
We turn to decays of the $Z$ into 3 jets and 2 jets plus one photon.
At the parton level this means
\bear
\label{10} Z &\rightarrow& b\bar{b}G\;, \\
\label{11} Z &\rightarrow& b\bar{b}\gamma\;.
\ear
We have studied these decays extensively in \cite{1}. In this note we
present some further calculations of CP--odd quantities for
(\ref{10}), (\ref{11}) and estimates of the number of events which
are necessary to detect CP--violating couplings of a magnitude given
by (\ref{8}). We will describe now various analysis scenarios.
\bigskip
\subsubsection*{I \underline{Analysis of $Z\rightarrow$ 3 jets,
               flavour blind case}}
Consider the decay $Z\rightarrow$ 3 jets (\ref{10}):
\be\label{12}
Z \rightarrow {\rm jet}(k_1) + {\rm jet}(k_2) + {\rm jet}(k_3) \;.
\ee
Let us assume that one can select events tagged by at least one $B$
decay and that the jets are ordered according to the magnitude of
their momenta (cf.\ (3.26a) of \cite{1}):
\be\label{13}
|\vec{k}_1| \ge |\vec{k}_2| \ge|\vec{k}_3| \;.
\ee
This the ``flavour blind'' case considered in \cite{1}, where we
showed that in the limit $m_b=0$ only the couplings $\hat{h}_{Vb}$,
$\hat{h}_{Ab}$ can induce CP--violating effects in 3 jet decays. We
found furthermore that all such CP--odd effects are then
proportional to the following linear combination of
$\hat{h}_{Vb}$, $\hat{h}_{Ab}$:
\be\label{14}
\hat{h}_{b}:=\hat{h}_{Ab}g_{Vb}-\hat{h}_{Vb}g_{Ab} \;,
\ee
where $g_{Vb}$, $g_{Ab}$ are the SM vector and axial vector
couplings of the $b$ quark to the $Z$ boson:
\be\label{15}
g_{Vb}=-\frac{1}{2}+\frac{2}{3}\sin^2\vartheta_W\;,\quad
g_{Ab}=-\frac{1}{2} \;.
\ee
{}From the limit on $\hat{h}^2_{Vb}+\hat{h}^2_{Ab}$ given in (\ref{8})
we get
\be\label{151}
|\hat{h}_{b}|\le 1.2 \;.
\ee
Assuming now that only $\hat{h}_{Vb}$, $\hat{h}_{Ab}$ are nonzero we
calculate the expectation values and the variances of the following
CP--odd observables (cf.\ (3.39) of \cite{1} and (2.20) of \cite{3}):
\be\label{16}
T^{(a)}_{ij}=\hat{k}_{ai}\hat{n}_j
+\hat{k}_{aj}\hat{n}_i \quad (a=1,2,3) \;.
\ee
Here $i,j$ ($1\le i,j \le 3$) are the Cartesian vector indices
(with 3 being the positron beam direction) and
\be\label{17}
\hat{\vec k}_{a}=\frac{{\vec k}_a}{|{\vec k}_a|}\;,\quad
\hat{\vec n}=\frac{\hat{\vec k}_{1}\times \hat{\vec k}_{2}}
{|\hat{\vec k}_{1}\times \hat{\vec k}_{2}|} \;.
\ee
We use a $y$-cut to define our sample of 3 jet events at parton
level, i.~e.\ we require
\be\label{18}
\frac{(k_a+k_b)^2}{m_Z^2}\ge\yc\;,\quad (1\le a\neq b\le 3)\;,
\ee
where $\yc$ is a conveniently chosen parameter. For a given $\yc$
we can expand the width and the expectation values of the
correlations $T^{(a)}_{33}$ as follows:
\bear
\Gamma(Z\rightarrow b\bar{b}G) &=& \Gamma^{\rm SM}_{b\bar{b}G}+
\left[(\hat{h}_{Vb})^2+(\hat{h}_{Ab})^2\right]
\Gamma^{\prime}_{b\bar{b}G} \;,\label{19}\\[2mm]
\langle T^{(a)}_{33} \rangle \Gamma(Z\rightarrow b\bar{b}G) &=&
\Gamma^{\rm SM}_{b\bar{b}G}Y^{(a)}\hat{h}_{b}
\quad (a=1,2,3)\;. \label{20}
\ear
Numerical results for $\Gamma^{\rm SM}_{b\bar{b}G}$,
$\Gamma'_{b\bar{b}G}$ and $Y^{(a)}$ are presented in Tables 2,3 for
$\yc$=0.03, 0.05 and 0.10. We also list the values for the variances
$\langle(T^{(a)}_{33})^2\rangle$ as calculated with the SM amplitude
alone. With this information we can calculate the number $\Nc$ of
events $Z\rightarrow b\bar{b}G$ within the corresponding $y$-cuts
which are needed in order to see a nonzero expectation value for
$T^{(a)}_{33}$ with a significance of 1 s.~d., given a value for
$\hat{h}_{b}$. We also list the corresponding total number $\Nt$
of $Z$ bosons needed. We have for a measurement of $T^{(a)}_{33}$:
\par
\hfill\parbox{13.4cm}{
\begin{eqnarray*}
\Nc & = & \frac{1}{|\hat{h}_{b}|^2}
\frac{\langle(T^{(a)}_{33})^2\rangle}{|Y_a|^2} \;, \\
\Nt &=& \Nc \frac{\Gamma_Z}{\Gamma^{\rm SM}_{b\bar{b}G}} \;.
\end{eqnarray*} }
\hfill\parbox{0.8cm}{\bear\label{21}\ear }
\par\noindent
Here $\Gamma_Z$ is the total $Z$ width and we set any non-standard
contribution to the widths to zero. For the numerics we use
$\Gamma_Z=2497\;{\rm MeV}$ \cite{20}. We see from Table 3 that for
$|\hat{h}_{b}|=1$ -- which is perfectly allowed by the experimental
results for $R_b$ (cf.\ (\ref{151})) -- the number of $Z$ bosons
required to see CP--odd effects is not outside the reach of today's
experiments.
Note, however, that in calculating $\Nt$ (\ref{21}) we have assumed
all experimental efficiencies, for $B$-tagging etc., to be
equal to one. In real life, efficiencies less than one are
unavoidable and will increase the number of $Z$ bosons required
to see CP--odd effects of a given magnitude. This remark applies
as well to all numbers $\Nt$ given below for other observables.
\par
Now we turn to ``optimal'' observables
[21--23].
Let us assume
that only the distribution of the unit momenta $\hat{\vec k}_{a}$
(a=1,2,3) of the three jets is analysed. Then we write the decay
distribution of the 3 jets in the reaction
$e^+e^-\rightarrow Z \rightarrow b\bar{b}G
\rightarrow 3\;\,{\rm jets}$
in the following way (cf.\ (6) and (31) of \cite{18}):
\par
\hfill\parbox{13.4cm}{
\begin{eqnarray*}
\frac{1}{\Gamma(Z \rightarrow b\bar{b}G)}d\!\!\!\!\!&\Gamma&
\!\!\!\!\!(Z\rightarrow b\bar{b}G\rightarrow 3\;\,{\rm jets})\,=\\
&=& \frac{1}{\Gamma^{\rm SM}_{b\bar{b}G}}
\Big(S_0+S_1\hat{h}_{b}+\,{\rm terms\,\;quadratic\,\;in\,\;}
\hat{h}_{Vb}, \hat{h}_{Ab}\Big)d\phi\;.
\end{eqnarray*} }
\hfill\parbox{0.8cm}{\bear\label{22}\ear }
\par\noindent
Here
\be\label{23}
d\phi=\delta[(\hat{\vec k}_{1}\times \hat{\vec k}_{2})\!\cdot\!
\hat{\vec k}_{3}]d\Omega_1d\Omega_2d\Omega_3
\Theta(|\vec{k}_1|-|\vec{k}_2|)\Theta(|\vec{k}_2|-|\vec{k}_3|)
\ee
is the phase space measure with $d\Omega_a$ the solid angle element
to $\hat{\vec k}_{a}$ (a=1,2,3). The $\delta$-function in (\ref{23})
takes into account that a three body final state is planar.
A suitable observable to measure $\hat{h}_{b}$ is then
\be\label{24}
O=\frac{S_1}{S_0}\;.
\ee
This observable is the optimal one if in (\ref{22}) quadratic terms
in $\hat{h}_{Vb}$, $\hat{h}_{Ab}$ are negligible. The explicit form
of $S_0$, $S_1$ is easily derived from the formulae given in \cite{1}.
We obtain:
\par
\hfill\parbox{13.4cm}{
\begin{eqnarray*}
S_0 \!&=&\! (g^2_{Vb}+g^2_{Ab})
\frac{24\alpha_s\Gamma_{\nu_e\bar{\nu}_e}}{\pi^3}
\frac{x_1x_2x_3}{y_1y_2y_3}\left\{\left[ x_1^3\left(
1+(\hat{\vec p}_+\!\!\cdot\hat{\vec k}_1)^2\right)\right] +
\begin{array}{c}{\rm\scriptstyle cycl.\;\,perm.'s} \\[-1mm]
{\rm\scriptstyle in\;\,(1,2,3)} \end{array}
\right\} \,, \\[2mm]
S_1 \!&=&\! \frac{24\alpha_s\Gamma_{\nu_e\bar{\nu}_e}}{\pi^3}
x_1x_2x_3 \,\times\\
&& \qquad\quad \times
\bigg\{\bigg[x_1^2x_2
\bigg(\frac{1}{y_3}-\frac{1}{y_2}\bigg)
\,\hat{\vec p}_+\!\!\cdot
(\hat{\vec k}_1\times \hat{\vec k}_2) \,
\hat{\vec p}_+\!\!\cdot\hat{\vec k}_1 \bigg]+
\begin{array}{c}{\rm\scriptstyle cycl.\;\,perm.'s} \\[-1mm]
{\rm\scriptstyle in\;\,(1,2,3)} \end{array}
\;\bigg\}\;.
\end{eqnarray*} }
\hfill\parbox{0.8cm}{\bear\label{25}\ear }
\par\noindent
Here $\hat{\vec p}_+$ is the unit vector in the direction of the
$e^+$ beam and
\be\label{26}
x_a=\frac{k_{a0}}{m_Z}\;,\quad y_a=1-2x_a \quad (a=1,2,3)\;.
\ee
We have written $O$ in (\ref{24}), (\ref{25}) in such a way that it
is a perfect CP--odd observable also for 3 jet decays of the $Z$ as
observed experimentally where the 3 jets are in general not exactly
planar due to photon radiation in the initial and/or final state, jet
reconstruction errors etc. For the expectation value we write
\be\label{27}
\langle O \rangle \Gamma(Z\rightarrow b\bar{b}G) =
\Gamma^{\rm SM}_{b\bar{b}G} C \hat{h}_{b} \;.
\ee
In Table 4 we present the results of our numerical calculations
for $C$ for three values of $\yc$. The quantity $C$ also represents
the expectation value of $O^2$ in the SM (cf.\ \cite{18}):
\be\label{28}
\langle O^2 \rangle_{\rm SM}=\frac{\int S_0
(S_1/S_0)^2
d\phi}{\int S_0d\phi}=C\;.
\ee
For the numbers  $\Nc$ and $\Nt$ required to see an effect at the
1 s.~d.\ level we have here
\par
\hfill\parbox{13.4cm}{
\begin{eqnarray*}
\Nc & = & \frac{1}{|\hat{h}_{b}|^2} \frac{1}{C} \;, \\
\Nt &=& \Nc \frac{\Gamma_Z}{\Gamma^{\rm SM}_{b\bar{b}G}} \;.
\end{eqnarray*} }
\hfill\parbox{0.8cm}{\bear\label{29}\ear }
\par\noindent
These numbers are also listed in Table 4. Comparing the results of
Tables 3 and 4 we find that the optimal observable (\ref{27})
leads to some but not a dramatic gain in sensitivity compared to
the observables $T^{(a)}_{33}$.
\bigskip
\subsubsection*{II \underline{Analysis of $Z\rightarrow$ 3 jets,
   identification of the highest energy jet} \protect\\ \hphantom{II}
   \underline{as coming from $b$ or $\bar b$ fragmentation}}
Here we consider the following type of analysis: In the decay
$Z\rightarrow$ 3 jets at least one $B$ hadron is observed. When the
three jets are ordered according to (\ref{12}), (\ref{13}) one
requires that the jet 1 which has the highest absolute value of
momentum contains a $B$ hadron. Let us assume that after
this selection of events the further analysis proceeds as in I with
the jet ordering criterion (\ref{13}).
\par
No we discuss the implications of the selection of events as
described above at parton level. If we order the jets in
$Z\rightarrow b\bar{b}G$
according to (\ref{13}) we can distinguish the 6 classes of events
shown in Table 5. With the procedure described above we select only
the events corresponding to the first 4 classes in Table 5.
\par
The further analysis follows the same lines as in I. In Table 6 we
list  $\Gamma^{\rm SM}_{b\bar{b}G,{\rm II}}$ and
$\Gamma'_{b\bar{b}G,{\rm II}}$ defined as in (\ref{19}) but with the
selection II imposed. Comparing the results of Tables 2 and 6 we see
that the 4 subclasses account for over 90 \% of the events, i.e.\
it is very unlikely to have a gluon jet as most energetic jet.
In Table 7 we list the values for parameters related to the CP--odd
observables $T^{(a)}_{33}$ ($a=1,2,3$) of (\ref{16}) again with the
selection II imposed:
\be\label{291}
\langle T^{(a)}_{33} \rangle_{\rm II} \Gamma_{\rm II}(Z\rightarrow
b\bar{b}G) = \Gamma^{\rm SM}_{b\bar{b}G,{\rm II}} Y^{(a)}_{\rm II}
\hat{h}_{b} \;.
\ee
We discuss now the optimal observable for analysis II. We have here
\par
\hfill\parbox{13.4cm}{
\begin{eqnarray*}
\frac{1}{\Gamma_{\rm II}(Z \rightarrow b\bar{b}G)}d\!\!\!\!\!
&\Gamma_{\rm II}&\!\!\!\!\!
(Z\rightarrow b\bar{b}G\rightarrow 3\;\,{\rm jets})\,=\\
&=& \frac{1}{{\Gamma}^{\rm SM}_{b\bar{b}G,{\rm II}}}
\Big(\tilde{S}_0+\tilde{S}_1\hat{h}_{b}+\,
{\rm terms\,\;quadratic\,\;in\,\;}
\hat{h}_{Vb}, \hat{h}_{Ab}\Big)d\phi\;,
\end{eqnarray*} }
\hfill\parbox{0.8cm}{\bear\label{321}\ear }
\par\noindent
where
\par
\hfill\parbox{13.4cm}{
\begin{eqnarray*}
\tilde{S}_0 &=& (g^2_{Vb}+g^2_{Ab})
\frac{12\alpha_s\Gamma_{\nu_e\bar{\nu}_e}}{\pi^3}
x_1x_2x_3\bigg\{\bigg[ \frac{1}{y_1y_2}\bigg( x_1^2\left(1+
(\hat{\vec p}_+\!\!\cdot\hat{\vec k}_1)^2\right)+ \\
&& \qquad\qquad\qquad\qquad+x_2^2
\left(1+(\hat{\vec p}_+\!\!\cdot\hat{\vec k}_2)^2\right)\bigg)\bigg]
+ [2\leftrightarrow 3] \bigg\} \,, \\[2mm]
\tilde{S}_1 &=& \frac{24\alpha_s\Gamma_{\nu_e\bar{\nu}_e}}{\pi^3}
x_1x_2x_3\,\times \\
&& \qquad\times \bigg\{\bigg[
x_1x_2\,\hat{\vec p}_+\!\!\cdot
(\hat{\vec k}_1\times \hat{\vec k}_2)
\bigg( \frac{x_2}{y_1}\hat{\vec p}_+\!\!\cdot\hat{\vec k}_2-
\frac{x_1}{y_2}\hat{\vec p}_+\!\!\cdot\hat{\vec k}_1 \bigg) \bigg]
+[2\leftrightarrow 3]\;\bigg\}\;.
\end{eqnarray*} }
\hfill\parbox{0.8cm}{\bear\label{322}\ear }
\par\noindent
The optimal observable is then
\be\label{323}
O_{\rm II}=\frac{\tilde{S}_1}{\tilde{S}_0}\;,
\ee
and we write for its expectation value
\be\label{324}
\langle O_{\rm II} \rangle \Gamma_{\rm II}(Z\rightarrow b\bar{b}G) =
\Gamma^{\rm SM}_{b\bar{b}G,{\rm II}} \,C_{\rm II} \,\hat{h}_{b} \;.
\ee
Values for $C_{\rm II}$ and $\Nc$, $\Nt$ defined in analogy to
(\ref{29}) are listed in Table 8.
\bigskip
\subsubsection*{III \underline{Analysis of $Z\rightarrow$ 3 jets,
   identification of the second highest energy} \protect\\
   \hphantom{III}
   \underline{jet as coming from $b$ or $\bar b$ fragmentation}}
Again we consider decays $Z\rightarrow$ 3 jets where at least one
$B$ hadron is observed, with the jet ordering (\ref{13}).
Contrary to the analysis II we now demand a $B$ tag
in jet 2, which is the jet with the second highest energy. Looking
through Table 5 we see that this corresponds to the
event classes 1,2 and 5,6. Note that this selection III is not
complementary to the selection II above; events where the least
energetic jet comes from the fragmentation of a
gluon are used both for analysis II and III.
\par
In Table 9 we give the values
$\Gamma^{\rm SM}_{b\bar{b}G,{\rm III}}$ and
$\Gamma'_{b\bar{b}G,{\rm III}}$ (cf.\ (\ref{19})) for the
selection III. The definition of the numbers $Y^{(a)}_{\rm III}$
characterizing the expectation values of the CP--odd observable
$T^{(a)}_{33}$ in the event sample III is identical to (\ref{291}),
\be\label{90}
\langle T^{(a)}_{33} \rangle_{\rm III} \Gamma_{\rm III}(Z\rightarrow
b\bar{b}G) = \Gamma^{\rm SM}_{b\bar{b}G,{\rm III}}Y^{(a)}_{\rm III}
\hat{h}_{b} \;,
\ee
and we list the relevant results in Table 10.
\par
The optimal observable for the present analysis is
\be\label{91}
O_{\rm III}=\frac{\tilde{S}_1(1\leftrightarrow 2)}
{\tilde{S}_0(1\leftrightarrow 2)}\;,
\ee
where $\tilde{S}_0(1\leftrightarrow 2)$,
$\tilde{S}_1(1\leftrightarrow 2)$
are obtained from (\ref{322})
by exchanging the indices $(1\leftrightarrow 2)$.
The expectation value $\langle O_{\rm III} \rangle$ defines the
numbers $C_{\rm III}$:
\be\label{92}
\langle O_{\rm III} \rangle \Gamma_{\rm III}(Z\rightarrow b\bar{b}G)
=\Gamma^{\rm SM}_{b\bar{b}G,{\rm III}}\,C_{\rm III} \,\hat{h}_{b}\;.
\ee
In Table 11 we present values for $C_{\rm III}$ and $\Nc$, $\Nt$
for various values of $\yc$. It is clear why the analysis III
has a higher sensitivity compared to analysis II:
As noted above, it is very unlikely that jet 1 is a gluon jet.
If one requires jet 2 to come from the fragmentation of
a $b$ ($\bar b$), one has then with a high probability the
full information of the parton content of the jets.
\bigskip
\subsubsection*{IV \underline{Analysis of $Z\rightarrow$ 3 jets
   with flavour identification}}
Here we discuss the case where two of the three jets in
$Z\rightarrow$ 3 jets have been tagged by the observation of a $B$
hadron as coming from the fragmentation of a $b$ or $\bar b$ quark.
Knowledge of the charge of the parent quark of the jet is
\underline{not} required. Let $\hat{\vec k}_+$ ($\hat{\vec k}_-$)
be the momentum direction of the $\bar b$ ($b$) jet. Then we can
form the following CP--odd tensor observable:
\be\label{30}
T'_{ij}=(\hat{\vec k}_+ -\hat{\vec k}_-)_i\left(\frac{\hat{\vec k}_+
\!\times\hat{\vec k}_-}{|\hat{\vec k}_+ \!\times\hat{\vec k}_-|}
\right)_j+(i\leftrightarrow j)\;,
\ee
where $1\le i,j\le 3$. Note that $T'_{ij}$ is invariant under
$\hat{\vec k}_+ \leftrightarrow \hat{\vec k}_-$. Thus the correct
and the wrong assignment of the momenta $\hat{\vec k}_\pm$ to the
$b$ and $\bar b$ jets give the same result for $T'_{ij}$. As above
we compute
\be\label{31}
\langle T'_{33} \rangle \Gamma(Z\rightarrow b\bar{b}G) =
\Gamma^{\rm SM}_{b\bar{b}G}Y'\hat{h}_{b} \;.
\ee
The quantities $\Gamma(Z\rightarrow b\bar{b}G)$ and
$\Gamma^{\rm SM}_{b\bar{b}G}$ for given $\yc$ are here identical to
the ones defined in (\ref{19}) for analysis I. The corresponding
numerical results are given in Table 2. Numerical results for $Y'$
and some other parameters related to the observable $T'_{33}$ are
collected in Table 12. The numbers $\Nt$ required to see possible
CP--odd effects at the 1 s.d.\ level are surprisingly low here.
We have, however, to remember that in our calculations all
efficiencies were assumed to be equal to one. In reality, the
double $B$-tag efficiency is considerably smaller than the efficiency
for a single $B$-tag. Thus the gain in sensitivity which is in
principle obtainable by the double tag method IV may in practice be
partly or completely lost due to the smaller efficiency.
\bigskip
\subsubsection*{V \underline{Analysis of $Z\rightarrow$ 2 jets
               + $\gamma$}}
Here we discuss the decay (\ref{11}):
\be\label{32}
Z \rightarrow b\bar{b}\gamma \rightarrow
{\rm jet}(k_+) + {\rm jet}(k_-) + \gamma(k) \;.
\ee
Experimentally this class of events can be selected by requiring two
jets, at least one $B$ hadron tag and one isolated photon. To define
the event sample we use again the $y$-cut (\ref{18}) applied to the
two jets and the photon as third ``jet''. We write then for the
width similarly to (\ref{19}):
\be\label{33}
\Gamma(Z\rightarrow b\bar{b}\gamma)=\Gamma^{\rm SM}_{b\bar{b}\gamma}
+\left[(\hat{f}_{Vb})^2+(\hat{f}_{Ab})^2\right]
\Gamma^{\prime}_{b\bar{b}\gamma} \;.
\ee
Numerical results for $\Gamma^{\rm SM}_{b\bar{b}\gamma}$ and
$\Gamma^{\prime}_{b\bar{b}\gamma}$ are shown in Table 13.
\par
As CP--odd observable we can choose here again $T'_{ij}$ defined in
(\ref{30}) where the assignment of the momenta $k_+$ and $k_-$
to the $b$ and $\bar b$ jet or vice versa does not matter. The
expectation values of $T'_{ij}$ depend only on $\hat{f}_{b}$
defined as
\be\label{34}
\hat{f}_{b}:=\hat{f}_{Ab}g_{Vb}-\hat{f}_{Vb}g_{Ab} \;.
\ee
This can be seen from the formulae in Appendix B of \cite{1}. The
restriction on $(\hat{f}_{Vb})^2+(\hat{f}_{Ab})^2$ in (\ref{8})
leads to
\be\label{35}
|\hat{f}_{b}|\le16.7\;.
\ee
For the expectation value of $T'_{33}$ we get here:
\be\label{36}
\langle T'_{33} \rangle \Gamma(Z\rightarrow b\bar{b}\gamma) =
\Gamma^{\rm SM}_{b\bar{b}\gamma}Y'_\gamma\hat{f}_{b} \;.
\ee
Numerical results for $Y'_\gamma$ and for
$\langle (T'_{33})^2 \rangle$, $\Nc$ and $\Nt$ defined as
\par
\hfill\parbox{13.4cm}{
\begin{eqnarray*}
\Nc & = & \frac{1}{|\hat{f}_{b}|^2}
\frac{\langle(T'_{33})^2\rangle}{|Y'_\gamma|^2} \;, \\
\Nt &=& \Nc \frac{\Gamma_Z}{\Gamma^{\rm SM}_{b\bar{b}\gamma}}
\end{eqnarray*} }
\hfill\parbox{0.8cm}{\bear\label{37}\ear }
\par\noindent
are shown in Table 14. Here the same procedure for the calculation
as explained after (\ref{20}) was followed.
\par
The decay $Z \rightarrow b\bar{b}\gamma$ was studied with respect
to CP--violating effects also by Abraham and Lampe \cite{7}. They
used various CP--odd asymmetries.
\par
\bigskip
Finally we make some concluding remarks. In this paper we have
studied the hypothesis that CP--violating couplings are responsible
for a small increase in the width $\Gamma(Z \rightarrow b\bar{b}X)$
compared to the SM prediction. We have shown that this can be checked
by direct searches for CP violation using appropriate CP--odd
observables. The number of $Z$ bosons required to obtain significant
information on the relevant parameters of CP violation is well
within the reach of present LEP experiments. We should emphasize
here that the bounds (\ref{8}) obtained for the CP--violating
parameters from the width $\Gamma(Z \rightarrow b\bar{b}X)$
depend crucially on the ansatz (\ref{3}) for the amplitude $\cal M$.
If new couplings play a role in the decays $Z \rightarrow b\bar{b}X$
one would in general expect that they
contribute both to the CP--conserving and
the CP--violating part of the amplitude $\cal M$. Then the
CP--conserving new couplings may reduce the width.
This can happen for
instance in extensions of the SM with several Higgs
doublets \cite{22}. For such models the width measurement,
i.e.\ the ratio (\ref{1})
yields a margin for CP--violating couplings
which is bigger than the bounds (\ref{8}) suggest.
Clearly, the width measurement
is no substitute for a direct search for CP violation. In \cite{1}
we have, in fact, given a \underline{complete} list of CP relations
which can be checked in three body decays of the $Z$ (cf.~Table 1
and (3.12), (3.28) of \cite{1}). In the present paper we have
considered tensor and optimal observables. The latter turn out to
have very small variances in theory. In real life, resolution
effects, measurement errors etc.\ will certainly increase the
variances and one has to check in each case if these observables
are then still better than e.g.\ the $T_{33}$ variables. Looking
through the numbers of the Tables we see that the most promising
way to search for CP--violating effects is open if two jets can be
tagged as coming from a $b$ or $\bar b$ quark (analyses IV, V). If
also the charge of the jets can be determined -- for instance by
observing the charge of the lepton in a semileptonic $B$ decay --
one has further interesting CP--odd observables at one's disposal
(cf.\ \cite{1}).
\par
In our opinion a study of CP--violating couplings in $Z$ decays to
final states containing $B$ hadrons deserves attention. We emphasize
again that we are concerned here not with CP violation in $B$ hadron
decays but with CP violation in the $Zb\bar b$, $Zb\bar bG$ and
$Zb\bar b\gamma$ vertices. We encourage experimentalists to
explore this field.
\bigskip
\section*{Acknowledgments}
The authors would like to thank P.~Bock, W.~Hollik, J.~von Krogh,
M.~Steiert, and M.~Wunsch for discussions and for providing useful
information. We are especially grateful to J.~von Krogh for the
discussions which led to the investigation of the analysis
procedures II and III.
\newpage
\section*{Table Captions}
\begin{description}
\item[Table 1:] \quad Contribution to the width for
   $Z\rightarrow b\bar{b}X$ from different anomalous CP--violating
   couplings. The first column lists the coupling parameter, the
   second column the final state to which the coupling contributes.
   In the third column we give the results for $\Delta\Gamma$
   where the $b$ quark mass has been set to zero.
\item[Table 2:] \quad The SM width for $Z\rightarrow b\bar{b}G$ and
   the width $\Gamma^\prime_{b\bar{b}G}$ as defined in (\ref{19})
   for three values of $\yc$ (\ref{18}).
\item[Table 3:] \quad Values for parameters related to the CP--odd
   observables $T^{(a)}_{33}$ ($a$=1,2,3) defined in (\ref{16}).
   Listed are $\yc$, $Y^{(a)}$, $\langle(T^{(a)}_{33})^2\rangle$,
   $|\hat{h}_b|^2\Nc$ and $|\hat{h}_b|^2\Nt$ (cf.\ (\ref{18}),
   (\ref{20}), (\ref{21})).
\item[Table 4:] \quad Values for parameters related to the CP--odd
   optimal observable $O$ defined in (\ref{24}), (\ref{25}).
   Listed  are $\yc$ (\ref{18}), $C$ (\ref{27}), (\ref{28}),
   $\Nc$ and $\Nt$ (\ref{29}).
\item[Table 5:] \quad The 6 possibilities for $b$, $\bar{b}$ and $G$
   to give 3 jets with momenta
   $|\vec{k}_1| \ge |\vec{k}_2| \ge |\vec{k}_3|$.
   Only the events corresponding to the
   first 4 rows satisfy the selection criterion II.
\item[Table 6:] \quad The SM width for $Z\rightarrow b\bar{b}G$ and
  the width $\Gamma^\prime_{b\bar{b}G,{\rm II}}$ as defined in
  (\ref{19}) for three values of $\yc$ (\ref{18}) and selection
  II imposed.
\item[Table 7:] \quad Values for parameters related to the CP--odd
   observables $T^{(a)}_{33}$ ($a$=1,2,3) defined in (\ref{16}) but
   now with the selection II imposed. Listed are $\yc$, $Y^{(a)}$,
   $\langle(T^{(a)}_{33})^2\rangle$, $|\hat{h}_b|^2\Nc$ and
   $|\hat{h}_b|^2\Nt$ (cf.\ (\ref{18}), (\ref{20}), (\ref{21})).
\item[Table 8:] \quad Values for parameters related to the CP--odd
   optimal observable $O_{\rm II}$ relevant if the selection II is
   applied (cf.\ (\ref{321})--(\ref{324})).
\item[Table 9:] \quad The SM width
   $\Gamma^{\rm SM}_{b\bar{b}G,{\rm III}}$
   and the width $\Gamma^\prime_{b\bar{b}G,{\rm III}}$
   (cf.\ (\ref{19})) for selection III.
\item[Table 10:] \quad\hspace{-3mm}
   Values for parameters related to the CP--odd
   observables $T^{(a)}_{33}$ ($a$=1,2,3) for analysis III. Listed
   are $\yc$, $Y^{(a)}$, $\langle(T^{(a)}_{33})^2\rangle$,
   $|\hat{h}_b|^2\Nc$ and $|\hat{h}_b|^2\Nt$.
\item[Table 11:] \quad\hspace{-3mm}
   $C_{\rm III}$, $|\hat{h}_b|^2\Nc$ and $|\hat{h}_b|^2\Nt$
   for the optimal observable $O_{\rm III}$ of analysis III
   (cf.\ (\ref{91}), (\ref{92})) for three values of $\yc$.
\item[Table 12:] \quad \hspace{-3mm}
   Values for parameters related to the CP--odd
   observable $T^\prime_{33}$ (\ref{30}). Listed are $\yc$,
   $Y^\prime$, $\langle(T^\prime_{33})^2\rangle$,
   $|\hat{h}_b|^2\Nc$, $|\hat{h}_b|^2\Nt$.
\item[Table 13:] \quad \hspace{-3mm}
   Numerical results for $\Gamma^{\rm SM}_{b\bar{b}\gamma}$
   and $\Gamma^\prime_{b\bar{b}\gamma}$ as defined in (\ref{33}).
\item[Table 14:] \quad \hspace{-3mm}
   Values for parameters related to the CP--odd
   observable $T^\prime_{33}$ defined in (\ref{30}) but applied
   to the decay $Z\rightarrow b\bar{b}\gamma$ (\ref{32}). Listed
   are $Y^\prime_\gamma$, $\langle(T^\prime_{33})^2\rangle$,
   $|\hat{f}_b|^2\Nc$ and $|\hat{f}_b|^2\Nt$ for three values of
   $\yc$ (cf.\ (\ref{36}), (\ref{37}), (\ref{18})).
\end{description}
\newpage
\renewcommand{\arraystretch}{1.5}
\vspace*{2.0cm}
\centerline{{\large\bf Table 1}}
\vspace{0.4cm}
\begin{center}
\begin{tabular}{|c|c|c|} \hline
\rule[-3mm]{0mm}{8mm}
\hbox to 2.5cm{\hfil $\begin{array}{c}{\rm coupling}\\[-2mm]
{\rm parameter} \end{array}$ \hfil}  &
\hbox to 2.5cm{\hfil final state\hfil}  &
$\Delta\Gamma(Z\rightarrow b\bar{b}X)$  \\ \hline\hline
\rule[-10mm]{0mm}{23mm}
$\tilde{d}_b$ & $b\bar{b}$ &
$\begin{array}{l} |\tilde{d}_b|^2
\frac{\textstyle m_Z^3}{\textstyle 8\pi}= \\
\quad =(|\tilde{d}_b|\cdot 10^{17}{\rm e}^{-1}{\rm cm}^{-1})^2
\cdot 0.71 \;{\rm MeV} \end{array}$ \\  \hline
\rule[-10mm]{0mm}{23mm}
$\hat{f}_{Vb}$, $\hat{f}_{Ab}$ & $b\bar{b}\gamma$ &
$\begin{array}{l} \left[(\hat{f}_{Vb})^2+(\hat{f}_{Ab})^2\right]
\frac{\textstyle \alpha}{\textstyle 30\pi}\Gamma_{\nu_e\bar{\nu}_e}=
\\ \quad=\left[(\hat{f}_{Vb})^2+(\hat{f}_{Ab})^2\right]
\cdot 1.4\!\cdot\!10^{-2} \;{\rm MeV} \end{array}$ \\  \hline
\rule[-10mm]{0mm}{23mm}
$\hat{h}_{Vb}$, $\hat{h}_{Ab}$ & $b\bar{b}G$ &
$\begin{array}{l} \left[(\hat{h}_{Vb})^2+(\hat{h}_{Ab})^2\right]
\frac{\textstyle 2\alpha_s}{\textstyle 5\pi}\Gamma_{\nu_e\bar{\nu}_e}
= \\ \quad=\left[(\hat{h}_{Vb})^2+(\hat{h}_{Ab})^2\right]
\cdot 2.54 \;{\rm MeV} \end{array}$ \\  \hline
\end{tabular}
\end{center}
\vspace{1.4cm}
\centerline{{\large\bf Table 2}}
\vspace{0.4cm}
\begin{center}
\begin{tabular}{|@{\hspace{5mm}}c@{\hspace{4mm}}||c|c|} \hline
\rule[-3mm]{0mm}{8mm}
$y_{\rm cut}$ & $\Gamma^{\rm SM}_{b\bar{b}G}$ [MeV]
& $\Gamma^{\prime}_{b\bar{b}G}$ [MeV] \\ \hline\hline
0.03 & 127.82 & 2.214  \\  \hline
0.05 &  81.33 & 1.990  \\  \hline
0.10 &  34.69 & 1.437  \\  \hline
\end{tabular}
\end{center}
\newpage
\centerline{{\large\bf Table 3}}
\vspace{0.4cm}
\begin{center}
\begin{tabular}{|@{\hspace{5mm}}c@{\hspace{4mm}}||c|c|c|c|} \hline
 $y_{\rm cut}$ & $Y^{(1)}$ &
$\langle(T^{(1)}_{33})^2\rangle$
& $|\hat{h}_b|^2\,N_{\rm cut}$
 & $|\hat{h}_b|^2\,N_{\rm tot}$ \\ \hline\hline
0.03 & --0.00364 & 0.282 & 21 316 & 415 712 \\  \hline
0.05 & --0.00452 & 0.281 & 13 731 & 421 736 \\  \hline
0.10 & --0.00567 & 0.277 &  8 595 & 618 793 \\  \hline\hline
 $y_{\rm cut}$ & $Y^{(2)}$ &
$\langle(T^{(2)}_{33})^2\rangle$ & $|\hat{h}_b|^2\,N_{\rm cut}$
 & $|\hat{h}_b|^2\,N_{\rm tot}$ \\ \hline\hline
0.03 &  0.00445 & 0.280 & 14 083 & 274 654 \\  \hline
0.05 &  0.00565 & 0.277 &  8 677 & 266 511 \\  \hline
0.10 &  0.00737 & 0.272 &  5 002 & 360 148 \\  \hline\hline
 $y_{\rm cut}$ & $Y^{(3)}$ &
$\langle(T^{(3)}_{33})^2\rangle$ & $|\hat{h}_b|^2\,N_{\rm cut}$
 & $|\hat{h}_b|^2\,N_{\rm tot}$ \\ \hline\hline
0.03 &  0.00010  & 0.246 & 24 829 426 & 484 221 122 \\  \hline
0.05 & --0.00026 & 0.244 &  3 670 817 & 112 740 730 \\  \hline
0.10 & --0.00110 & 0.243 &    199 527 &  14 363 713 \\  \hline
\end{tabular}
\end{center}
\vspace{0.5cm}
\centerline{{\large\bf Table 4}}
\vspace{0.4cm}
\begin{center}
\begin{tabular}{|@{\hspace{5mm}}c@{\hspace{4mm}}||c|c|c|}
\hline
 $y_{\rm cut}$ & $C$  & $|\hat{h}_b|^2\,N_{\rm cut}$
 & $|\hat{h}_b|^2\,N_{\rm tot}$ \\ \hline\hline
0.03 & 0.0001113 & 8 988 & 175 332 \\  \hline
0.05 & 0.0001566 & 6 385 & 196 102 \\  \hline
0.10 & 0.0002382 & 4 197 & 302 314 \\  \hline
\end{tabular}
\end{center}
\vspace{0.5cm}
\centerline{{\large\bf Table 5}}
\vspace{0.4cm}
\begin{center}
\begin{tabular}{|c|c|c|} \hline
\rule[-3mm]{0mm}{8mm}
\hbox to 2cm{\hfil jet 1 \hfil} &
\hbox to 2cm{\hfil jet 2 \hfil} &
\hbox to 2cm{\hfil jet 3 \hfil} \\ \hline\hline
   $b$    & $\bar{b}$ &    $G$    \\  \hline
$\bar{b}$ &    $b$    &    $G$    \\  \hline
   $b$    &    $G$    & $\bar{b}$ \\  \hline
$\bar{b}$ &    $G$    &    $b$    \\  \hline
   $G$    &    $b$    & $\bar{b}$ \\  \hline
   $G$    & $\bar{b}$ &    $b$    \\  \hline
\end{tabular}
\end{center}
\newpage
\centerline{{\large\bf Table 6}}
\vspace{0.4cm}
\begin{center}
\begin{tabular}{|@{\hspace{5mm}}c@{\hspace{4mm}}||c|c|} \hline
\rule[-3mm]{0mm}{8mm}
$y_{\rm cut}$ & $\Gamma^{\rm SM}_{b\bar{b}G,{\rm II}}$ [MeV]
& $\Gamma^{\prime}_{b\bar{b}G,{\rm II}}$ [MeV] \\ \hline\hline
0.03 & 121.35 & 1.033  \\  \hline
0.05 &  75.63 & 0.963  \\  \hline
0.10 &  30.75 & 0.751  \\  \hline
\end{tabular}
\end{center}
\vspace{1.4cm}
\centerline{{\large\bf Table 7}}
\vspace{0.4cm}
\begin{center}
\begin{tabular}{|@{\hspace{5mm}}c@{\hspace{4mm}}||c|c|c|c|} \hline
 $y_{\rm cut}$ & $Y_{\rm II}^{(1)}$ &
$\langle(T^{(1)}_{33})^2\rangle_{\rm II}$
& $|\hat{h}_b|^2\,N_{\rm cut}$
 & $|\hat{h}_b|^2\,N_{\rm tot}$ \\ \hline\hline
0.03 & --0.00402 & 0.283 & 17 543 & 360 762 \\  \hline
0.05 & --0.00514 & 0.282 & 10 660 & 352 016 \\  \hline
0.10 & --0.00696 & 0.279 &  5 757 & 467 638 \\  \hline\hline
 $y_{\rm cut}$ & $Y_{\rm II}^{(2)}$ &
$\langle(T^{(2)}_{33})^2\rangle_{\rm II}$ &
$|\hat{h}_b|^2\,N_{\rm cut}$ &
$|\hat{h}_b|^2\,N_{\rm tot}$ \\ \hline\hline
0.03 &  0.00568 & 0.280 &  8 668 & 178 259 \\  \hline
0.05 &  0.00763 & 0.278 &  4 769 & 157 471 \\  \hline
0.10 &  0.01168 & 0.272 &  1 995 & 162 099 \\  \hline\hline
 $y_{\rm cut}$ & $Y_{\rm II}^{(3)}$ &
$\langle(T^{(3)}_{33})^2\rangle_{\rm II}$ &
$|\hat{h}_b|^2\,N_{\rm cut}$ &
$|\hat{h}_b|^2\,N_{\rm tot}$ \\ \hline\hline
0.03 & --0.00106 & 0.246 & 218 297 & 4 488 952 \\  \hline
0.05 & --0.00210 & 0.244 &  55 367 & 1 828 200 \\  \hline
0.10 & --0.00509 & 0.243 &   9 346 &   759 182 \\  \hline
\end{tabular}
\end{center}
\vspace{1.4cm}
\centerline{{\large\bf Table 8}}
\vspace{0.4cm}
\begin{center}
\begin{tabular}{|@{\hspace{5mm}}c@{\hspace{4mm}}||c|c|c|}
\hline
 $y_{\rm cut}$ & $C_{\rm II}$  & $|\hat{h}_b|^2\,N_{\rm cut}$
 & $|\hat{h}_b|^2\,N_{\rm tot}$ \\ \hline\hline
0.03 & 0.000244 & 4 106 & 84 494 \\  \hline
0.05 & 0.000368 & 2 714 & 89 552 \\  \hline
0.10 & 0.000717 & 1 394 & 113 252 \\  \hline
\end{tabular}
\end{center}
\newpage
\centerline{{\large\bf Table 9}}
\vspace{0.4cm}
\begin{center}
\begin{tabular}{|@{\hspace{5mm}}c@{\hspace{4mm}}||c|c|} \hline
\rule[-3mm]{0mm}{8mm}
$y_{\rm cut}$ & $\Gamma^{\rm SM}_{b\bar{b}G,{\rm III}}$ [MeV]
& $\Gamma^{\prime}_{b\bar{b}G,{\rm III}}$ [MeV] \\ \hline\hline
0.03 & 105.16 & 1.552  \\  \hline
0.05 &  65.07 & 1.378  \\  \hline
0.10 &  26.41 & 0.971  \\  \hline
\end{tabular}
\end{center}
\vspace{1.4cm}
\centerline{{\large\bf Table 10}}
\vspace{0.4cm}
\begin{center}
\begin{tabular}{|@{\hspace{5mm}}c@{\hspace{4mm}}||c|c|c|c|} \hline
 $y_{\rm cut}$ & $Y^{(1)}_{\rm III}$ &
$\langle(T^{(1)}_{33})^2\rangle_{\rm III}$
& $|\hat{h}_b|^2\,N_{\rm cut}$
 & $|\hat{h}_b|^2\,N_{\rm tot}$ \\ \hline\hline
0.03 & --0.00799 & 0.282 & 4 426 & 105 148 \\  \hline
0.05 & --0.01049 & 0.281 & 2 552 &  97 961 \\  \hline
0.10 & --0.01532 & 0.277 & 1 180 & 111 609 \\  \hline\hline
 $y_{\rm cut}$ & $Y^{(2)}_{\rm III}$ &
$\langle(T^{(2)}_{33})^2\rangle_{\rm III}$
 & $|\hat{h}_b|^2\,N_{\rm cut}$
 & $|\hat{h}_b|^2\,N_{\rm tot}$ \\ \hline\hline
0.03 &  0.00730 & 0.282 & 5 287 & 125 614 \\  \hline
0.05 &  0.00935 & 0.280 & 3 200 & 122 848 \\  \hline
0.10 &  0.01257 & 0.276 & 1 747 & 165 246 \\  \hline\hline
 $y_{\rm cut}$ & $Y^{(3)}_{\rm III}$ &
$\langle(T^{(3)}_{33})^2\rangle_{\rm III}$
 & $|\hat{h}_b|^2\,N_{\rm cut}$
 & $|\hat{h}_b|^2\,N_{\rm tot}$ \\ \hline\hline
0.03 & 0.00425 & 0.243 & 13 445 & 319 414 \\  \hline
0.05 & 0.00533 & 0.241 &  8 478 & 325 438 \\  \hline
0.10 & 0.00766 & 0.239 &  4 068 & 384 834 \\  \hline
\end{tabular}
\end{center}
\vspace{1.4cm}
\centerline{{\large\bf Table 11}}
\vspace{0.4cm}
\begin{center}
\begin{tabular}{|@{\hspace{5mm}}c@{\hspace{4mm}}||c|c|c|}
\hline
 $y_{\rm cut}$ & $C_{\rm III}$  & $|\hat{h}_b|^2\,N_{\rm cut}$
 & $|\hat{h}_b|^2\,N_{\rm tot}$ \\ \hline\hline
0.03 & 0.000377 & 2 650 & 62 957 \\  \hline
0.05 & 0.000553 & 1 807 & 69 450 \\  \hline
0.10 & 0.000985 & 1 015 & 96 016 \\  \hline
\end{tabular}
\end{center}
\newpage
\centerline{{\large\bf Table 12}}
\vspace{0.4cm}
\begin{center}
\begin{tabular}{|@{\hspace{5mm}}c@{\hspace{4mm}}||c|c|c|c|}
\hline
 $y_{\rm cut}$ & $Y'$ & $\langle(T^\prime_{33})^2\rangle$ &
$|\hat{h}_b|^2\,N_{\rm cut}$
 & $|\hat{h}_b|^2\,N_{\rm tot}$ \\ \hline\hline
0.03 & --0.02202 & 1.040 & 2 146 & 41 893 \\  \hline
0.05 & --0.02910 & 1.009 & 1 191 & 36 577 \\  \hline
0.10 & --0.04400 & 0.947 &   489 & 35 216 \\  \hline
\end{tabular}
\end{center}
\vspace{1.4cm}
\centerline{{\large\bf Table 13}}
\vspace{0.4cm}
\begin{center}
\begin{tabular}{|@{\hspace{5mm}}c@{\hspace{4mm}}||c|c|} \hline
\rule[-3mm]{0mm}{8mm}
$y_{\rm cut}$ & $\Gamma^{\rm SM}_{b\bar{b}\gamma}$ [keV]
& $\Gamma^{\prime}_{b\bar{b}\gamma}$ [keV] \\ \hline\hline
0.03 & 688.8 & 11.92  \\  \hline
0.05 & 438.0 & 10.72  \\  \hline
0.10 & 186.7 &  7.73  \\  \hline
\end{tabular}
\end{center}
\vspace{1.4cm}
\centerline{{\large\bf Table 14}}
\vspace{0.4cm}
\begin{center}
\begin{tabular}{|@{\hspace{5mm}}c@{\hspace{4mm}}||c|c|c|c|}
\hline
$y_{\rm cut}$ & $Y'_\gamma$ & $\langle(T^\prime_{33})^2\rangle$ &
$|\hat{f}_b|^2\,N_{\rm cut}$
 & $|\hat{f}_b|^2\,N_{\rm tot}$ \\ \hline\hline
0.03 & 0.02202 & 1.040 & 2 146 & 7 782 182 \\  \hline
0.05 & 0.02910 & 1.009 & 1 191 & 6 794 696 \\  \hline
0.10 & 0.04400 & 0.947 &   489 & 6 541 781 \\  \hline
\end{tabular}
\end{center}

\newpage
\begin{thebibliography}{99}
\bibitem{1} W.~Bernreuther, U.~L\"ow, J.~P.~Ma, O.~Nachtmann:
   Z.\ Phys.\ {\bf C 43} (1989) 117
\bibitem{2} W.~Bernreuther, O.~Nachtmann:
   Phys.\ Rev.\ Lett.\ {\bf 63} (1989) 2787
\bibitem{3} J.~K\"orner, J.~P.~Ma, R.~M\"unch, O.~Nachtmann,
   R.~Sch\"opf:\\ Z.\ Phys.\ {\bf C 49} (1991) 447
\bibitem{4} W.~Bernreuther, G.~W.~Botz, O.~Nachtmann, P.~Overmann:\\
   Z.\ Phys.\ {\bf C 52} (1991) 567
\bibitem{5} W.~Bernreuther, O.~Nachtmann:
   Phys.\ Lett.\ {\bf B 268} (1991) 424
\bibitem{6} W.~Bernreuther, O.~Nachtmann, P.~Overmann:
   Phys.\ Rev.\ {\bf D 48} (1993) 78
\bibitem{7} L.~Stodolsky: Phys.\ Lett.\ {\bf B 150} (1985) 221;\\
   F.~Hoogeveen, L.~Stodolsky: Phys.\ Lett.\ {\bf B 212}
   (1988) 505;\\
   J.~F.~Donoghue, B.~R.~Holstein, G.~Valencia:\\
   Int.\ J.\ Mod.\ Phys.\ {\bf A 2} (1987) 319;\\
   J.~F.~Donoghue, G.~Valencia:
   Phys.\ Rev.\ Lett.\ {\bf 58} (1987) 451;\\
   J.~Bernab\'eu, N.~Rius: Phys.\ Lett.\ {\bf B 232} (1989) 127;\\
   J.~Bernab\'eu, N.~Rius, A.~Pich:
   Phys.\ Lett.\ {\bf B 257} (1991) 219;\\
   M.~B.~Gavela, F.~Iddir, A.~Le Yaouanc, L.~Olivier, O.~P\`ene,
   J.~C.~Raynal:\\ Phys.\ Rev.\ {\bf D 39} (1989) 1870;\\
   A.~De Rujula, M.~B.~Gavela, O.~P\`ene, F.~J.~Vegas:\\
   Nucl.\ Phys.\ {\bf B 357} (1991) 311;\\
   S.~Goozovat, C.~A.~Nelson: Phys.\ Lett.\ {\bf B 267}
   (1991) 128;\\ Phys.\ Rev.\ {\bf D 44} (1991) 2818;\\
   G.~Valencia, A.~Soni: Phys.\ Lett.\ {\bf B 263} (1991) 517;\\
   K.~J.~Abraham, B.~Lampe: Phys.\ Lett.\ {\bf B 326} (1994) 175
\bibitem{8} P.~D.~Acton et al.\ (OPAL coll.):
   Phys.\ Lett.\ {\bf B 281} (1992) 405;\\
   D.~Buskulic et al.\ (ALEPH coll.):
   Phys.\ Lett.\ {\bf B 297} (1992) 459;\\
   R.~Akers et al.\ (OPAL coll.):
   preprint CERN--PPE--94--171 (1994)
\bibitem{9} R.~Jones: Talk given at the XXVII Int.\ Conf.\ on High
   Energy Physics, \\ Glasgow (July 1994)
\bibitem{9a} R.~Akers et al.\ (OPAL coll.):
   preprint CERN--PPE--94--106 (1994) \\
   (submitted to Z.\ Phys.\ {\bf C})
\bibitem{10} F.~Abe et al.\ (CDF coll.):
   Phys.\ Rev.\ {\bf D 50} (1994) 2966;\\
   Phys.\ Rev.\ Lett.\ {\bf 73} (1994) 225;
\bibitem{10a} A.~A.~Akhundov, D.~Yu.~Bardin, T.~Riemann:
   Nucl.\ Phys.\ {\bf B 276} (1986) 1;\\
   W.~Beenakker, W.~Hollik: Z.\ Phys.\ {\bf C 40} (1988) 141;\\
   J.~Bernab\'eu, A.~Pich, A.~Santamaria:
   Phys.\ Lett.\ {\bf B 200} (1988) 569;\\
   J.~Fleischer, O.~V.~Tarasov, F.~Jegerlehner:
   Phys.\ Lett.\ {\bf B 319} (1993) 249;\\
   M.~Vysotsky: Plenary talk given at the XXVII Int.\ Conf.\
   on High Energy Physics, Glasgow (July 1994)
\bibitem{10b} J.~Mnich: ``Precision tests of electroweak theory
   at LEP'', in Proc.\ $13^{\rm th}$ Int.\ Conf.\ Physics in
   Collision, Heidelberg 1993, eds.\ E.~E.~Kluge, K.~Tittel \\
   (Editions Fronti\`eres, Gif-sur-Yvette 1994)
\bibitem{11} B.~Ananthanarayan, S.~D.~Rindani:
   Phys.\ Rev.\ {\bf D 50} (1994) 4447
\bibitem{12} O.~Nachtmann: Nucl.\ Phys.\ {\bf B 127} (1977) 314
\bibitem{13} A.~V.~Efremov: Sov.\ J.\ Nucl.\ Phys.\ {\bf 28}
   (1978) 83
\bibitem{14} R.~H.~Dalitz, G.~R.~Goldstein, R.~Marshall:
   Z.\ Phys.\ {\bf C 42} (1989) 441
\bibitem{15} A.~V.~Efremov, L.~Mankiewicz, N.~A.~Tornquist:
   Phys.\ Lett.\ {\bf B 284} (1992) 394
\bibitem{16} P.~N.~Burrows: ``Studies of QCD $B$-Physics and Jet
   Handedness at SLD'', \\ XXXVIII Rencontres de Moriond
   ``QCD and High Energy Hadronic Interactions''
   (Editions Fronti\`eres, Gif-sur-Yvette 1993) \\
   H.~Masuda (SLD coll.): Talk given at the XXVII Int.\ Conf.\ on
   High Energy Physics, Glasgow (July 1994)
\bibitem{20} D.~Schaile: Plenary talk given at the XXVII Int.\
   Conf.\ on High Energy Physics, Glasgow (July 1994)
\bibitem{17} D.~Atwood, A.~Soni: Phys.\ Rev.\ {\bf D 45}
   (1992) 2405;\\ M.~Davier, L.~Duflot, F.~Le Diberder,
   A.~Roug\'e: Phys.\ Lett.\ {\bf B 306} (1993) 411
\bibitem{18} M.~Diehl, O.~Nachtmann: Z.\ Phys.\ {\bf C 62} (1994) 397
\bibitem{19} P.~Overmann: ``A new method to measure the $\tau$
   polarization at the $Z$ peak'', Dortmund university preprint
   DO--TH--93--24 (1993)
\bibitem{22} W.~Hollik: ``New physics from precision measurements'',
   \\ in: ``Polarization at LEP'', G.~Alexander et al.\ (eds.),
   CERN 88--06 (1988);\\
   G.~Girardi, W.~Hollik, C.~Verzegnassi:
   Phys.\ Lett.\ {\bf B 240} (1990) 492;\\
   W.~Hollik: private communication
\end{thebibliography}
\end{document}